\documentclass[12pt]{iopart}
\expandafter\let\csname equation*\endcsname\relax
\expandafter\let\csname endequation*\endcsname\relax 
\usepackage{iopams}
\usepackage{epsfig}
\usepackage{amssymb}
\usepackage{amsmath}
\usepackage{rotating}
\usepackage{latexsym}
\usepackage{graphicx}
\usepackage{subfig}
\usepackage{wrapfig}
\usepackage{listings}
\usepackage{float}
\usepackage{lipsum}
\usepackage{enumerate}
\usepackage{array}
\floatstyle{plain}
\newcommand{\hydrogen}{\textsuperscript{1}H}
\newcommand{\carbon}{\textsuperscript{13}C}

\newcommand{\Ham}{{\mathcal{H}}}
\newcommand{\kket}[1]{|{#1}\rangle}
\newcommand{\bbra}[1]{\langle{#1}|}

\newcommand{\ql}{\textquotedblleft}
\newcommand{\UU}{\hat{U}}
\newcommand{\VV}{\hat{V}}
\begin{document}
\title[Three path interference using nuclear magnetic resonance]{Three path interference using nuclear magnetic resonance: a test of the consistency of Born's rule}

\author{D K Park$\textsuperscript{1}$, O Moussa$\textsuperscript{1}$ and R Laflamme$\textsuperscript{1,2}$}

\address{$\textsuperscript{1}$Institute for Quantum Computing and Department of Physics, University of Waterloo, 200 University Ave W, Waterloo, Ontario N2L 3G1, Canada
}
\address{$\textsuperscript{2}$Perimeter Institute for Theoretical Physics, 31 Caroline St, Waterloo, Ontario N2L 2Y5, Canada
}
\ead{kpark@iqc.ca}

\begin{abstract}
The Born rule is at the foundation of quantum mechanics and transforms our classical way of understanding probabilities by predicting that interference occurs between pairs of independent paths of a single object. One consequence of the Born rule is that three way (or three paths) quantum interference does not exist. In order to test the consistency of the Born rule, we examine detection probabilities in three path intereference using an ensemble of spin-1/2 quantum registers in liquid state nuclear magnetic resonance (LSNMR). As a measure of the consistency, we evaluate the ratio of three way interference to two way interference. Our experiment bounded the ratio to the order of $10^{-3}\pm 10^{-3}$, and hence it is consistent with Born's rule.
\end{abstract}
\maketitle
\def\one{{\mathchoice {\rm 1\mskip-4mu l} {\rm 1\mskip-4mu l} {\rm
\mskip-4.5mu l} {\rm 1\mskip-5mu l}}}

\section{INTRODUCTION}
The Born rule is one of the fundamental postulates of quantum mechanics which states that if a quantum mechanical state is described by a wavefunction $\psi(\textbf{r},t)$, then the probability of finding a particle at \textbf{r} in the volume element $d^{3}r$ at time $t$ is~\cite{born}
\begin{equation}
p(\textbf{r},t)d^{3}r=\psi^{\ast}(\textbf{r},t)\psi(\textbf{r},t)d^{3}r=|\psi(\textbf{r},t)|^{2}d^{3}r.
\end{equation}
The Born rule has well described several experimental results, but no experiments (until the recent work by Sinha \textit{et al}.~\cite{sinha,sinha2}) have been performed to test directly the validity of this foundational theory of quantum mechanics. Thus, a deviation from the theory, if there is any, would not have been evident. Though quantum mechanics has been a very successful theory, it still does not fully satisfy our understanding of the universe. Therefore, it is very important to take steps towards experimental verification of the Born rule.

As a direct consequence of the Born rule, an interference pattern is produced when even a single particle travels through two slits. Quantum interference can be stated as a deviation from the classical interpretation of probability for mutually exclusive events (e.g. paths, slits, eigenstates and etc.)~\cite{sinha, sinha2}. For instance, quantum interference of two paths 1 and 2 is $\text{I}_{12}=\text{P}_{12}-(\text{P}_{1}+\text{P}_{2})$, where $\text{P}_{i}$ is the probability for a path configuration $i$. Similarly, quantum interference of three paths 1, 2 and 3 can be written as $\text{I}_{123}=\text{P}_{123}-(\text{P}_{12}+\text{P}_{13}+\text{P}_{23}-\text{P}_{1}-\text{P}_{2}-\text{P}_{3})$.
According to Born's rule, three paths probability is 
\begin{align}
\label{prob}
\text{P}_{123}&=|\psi_{1}+\psi_{2}+\psi_{3}|^{2} \nonumber \\
&=\text{P}_{1}+\text{P}_{2}+\text{P}_{3}+\text{I}_{12}+\text{I}_{13}+\text{I}_{23} \nonumber \\
&=\text{P}_{12}+\text{P}_{23}+\text{P}_{13}-\text{P}_{1}-\text{P}_{2}-\text{P}_{3}.
\end{align}
Therefore, the Born rule predicts that $\text{I}_{123}=0$~\cite{sorkin}. Here we introduce $\text{P}_{0}$ to denote the probability of detecting particles when all paths are blocked (ideally zero). Non-zero value of $\text{P}_{0}$ may rise in the actual implementation due to experimental errors such as detector noise. Thus, the measured quantity in the experiment is
\begin{equation}
\label{tint} \text{I}_{123}=\text{P}_{123}-\text{P}_{12}-\text{P}_{13}-\text{P}_{23}+\text{P}_{1}+\text{P}_{2}+\text{P}_{3}-\text{P}_{0}.
\end{equation}

The three path experiment tests whether \Eref{tint} is zero by observing the probabilities resulting from all possible combinations of independent paths being blocked and unblocked, and hence validate the Born rule. Note that this experiment is a more precise test for Born's rule than an experiment with two paths in which one has to measure the non zero interference pattern and compare it with the theoretical prediction~\cite{sinha}. We perform the experiment using an ensemble of spin-1/2s in an LSNMR quantum computer. This experiment does not only examines the Born rule directly, but also demonstrates the capability of LSNMR quantum computing for testing fundamental laws of quantum theory.

In this letter, we report the results of a three path experiment that exploits NMR quantum information processing (QIP) for testing the consistency of the Born rule. The remainder of the paper is organized as follows. Section 2 explains how to represent the three paths using energy eigenstates which can be implemented with NMR. Section 3 describes the experimental set up. The results of the experiment and the discussion of possible sources of errors follow in Section 4 and Section 5.

\section{ENCODING THREE PATHS IN ENERGY EIGENSTATES}
Here we describe how to translate the triple-slit experiment~\cite{sinha, sinha2} into a form that can be implemented with NMR. When a photon travels through one of the eight possible slit configurations, the initial single path (of the photon before arriving at the slits) state evolves into another state which is determined by the slit configuration. Some photons are lost (not detected) by arriving at a path that is blocked, and a superposition of the unblocked paths is created. Due to this loss of photons, the overall transformation can be described as a non-trace preserving map. We encode this non-trace preserving map by an implementable unitary transformation on a larger Hilbert space.

Consider a four-level system with energy eigenstates $\kket{0}\text{,}\ \kket{1}\text{,}\ \kket{2}\text{,}\ \kket{3}$. One can imagine the basis states $\kket{1}\text{,}\ \kket{2}\text{,}\ \kket{3}$ as encoding the path taken by a photon in the triple-slit experiment ~\cite{sinha, sinha2} as it travels through slit 1, 2 and 3, respectively. We can construct a superposition, $\kket{\psi^{\gamma}}$, of these four states to represent a particular slit pattern $\gamma$, as follows
\begin{equation}
\label{psib}
|\psi^{\gamma}\rangle=\beta|0\rangle+\sum^{3}_{k=1}\frac{\gamma_{k}}{\sqrt{3}}|k\rangle,
\end{equation}
and $\gamma$ is defined as
$$\gamma=\gamma_{1}\gamma_{2}\gamma_{3}\in\{000,001,010,100,110,101,001,111\},$$ where
\[\gamma_{k} = \left\{ 
\begin{array}{l l}
  0 & \quad \mbox{if path k is blocked}\\
  1 & \quad \mbox{if path k is unblocked}\\ \end{array} \right. , \]
and $\beta$ is determined from the normalization condition. The amplitude, $\beta$, of the state $\kket{0}$ captures the probability that a photon does not arrive at the detector due to any blocked paths. For example, when all three paths are open, $\beta$ = 0 and all photons reach the detector. On the other hand, for $\beta$ = 1, the state encoding the slit information is $\kket{0}$. This state represents that all three slits are closed and is used for calculating the background probability $\text{P}_0$ which can be non-zero due to experimental imperfections. \Tref{table} illustrates all possible slit patterns that match with the path configuration label $\gamma$, and the corresponding superposition states $\kket{\psi^{\gamma}}$.
\begin{table}[h]
\centering
\caption{Table of all possible path arrangements that can be formed from three independent paths and superposition states that encode each configuration. We first form the equal superposition state $\kket{\psi^{111}}$ to encode three slits, and then write other superposition states for the rest of slit configurations according to \Eref{psib} such that the amplitudes for each states $\kket{j}$ with $j\in\lbrace 1,\; 2,\; 3\rbrace$ are either 0 (when the path is blocked) or $\frac{1}{\sqrt{3}}$ (when the path is open) for all $\kket{\psi^{\gamma}}$, and the amplitude, $\beta$, of the state $\kket{0}$ is chosen to satisfy the normalization condition.\label{table}}
\epsfig{file=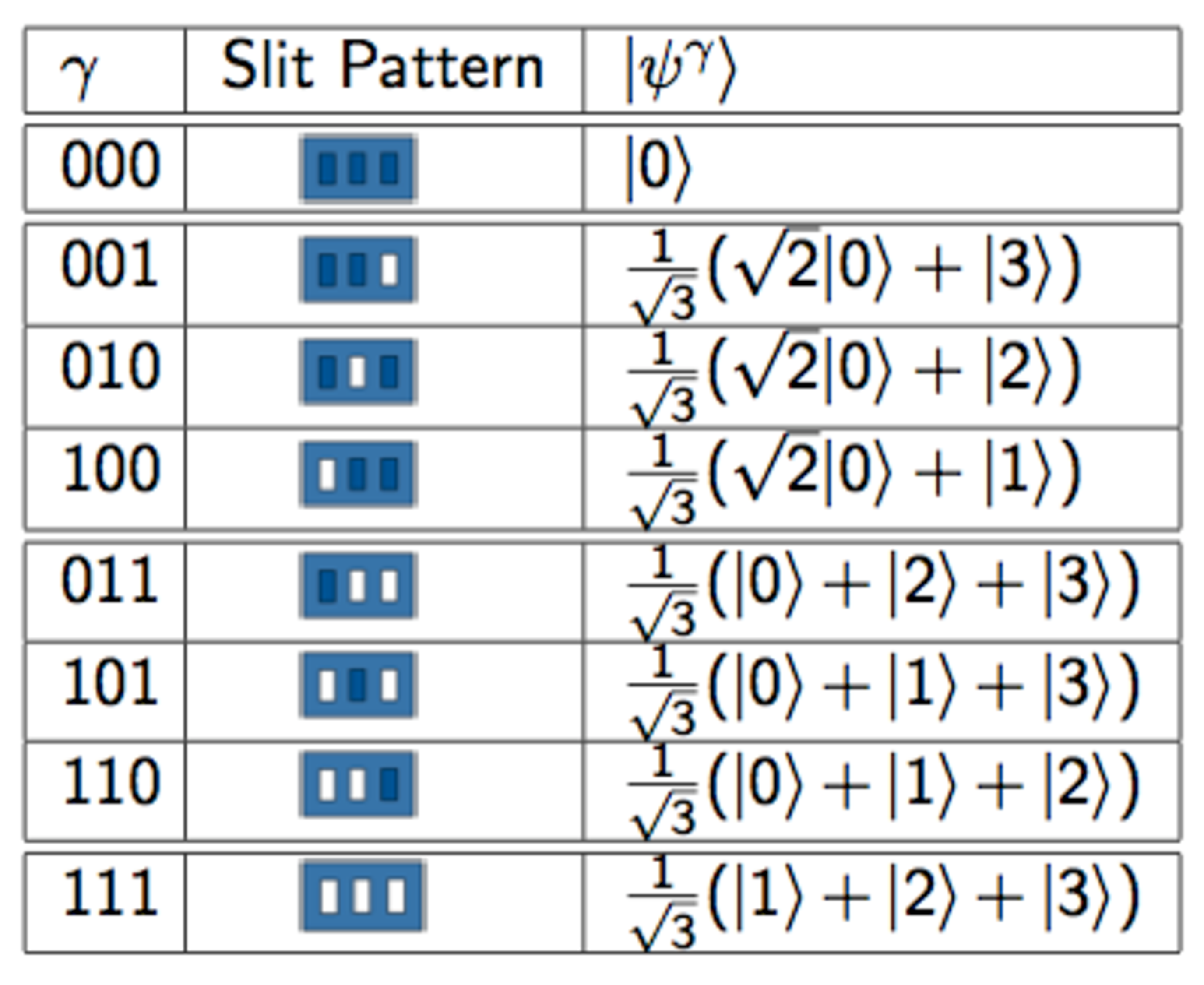,angle=270,width=0.4\textwidth}
\end{table}

Moreover, we introduce $\tau$ to parameterize the evolution between the coherence creation and detecting interference. $\tau$ is directly related to the position of the detector with respect to some central position in the triple-slit experiment~\cite{sinha, sinha2}. Suppose the state $\kket{\psi^{111}}\bbra{\psi^{111}}$ evolves under the Hamiltonian $\Ham_{0}=\text{E}_{0}|0\rangle\langle0|+\text{E}_{1}|1\rangle\langle1|+\text{E}_{2}|2\rangle\langle2|+\text{E}_{3}|3\rangle\langle3|$ for $\tau$. Then the evolved state would have the following form:
\begin{equation}
\label{evolved}
|\psi^{111}(\tau)\rangle=\sum^{3}_{j=1}\frac{e^{-i\triangle_{j}\tau}}{\sqrt{3}}|j\rangle,
\end{equation}
where $\triangle_{j}=\text{E}_{j}-\text{E}_{0}$. Born's rule for probability dictates that
\begin{equation}
\label{bornrule2}
\text{P}_{\gamma}(\tau)=|\langle\psi^{\gamma}|\psi^{111}(\tau)\rangle|^{2},
\end{equation}
where the subscript $\gamma$ indicates the path configuration of which the probability is measured.
We can analytically calculate $\text{P}_{\gamma}(\tau)$ for all $\gamma$ and confirm that \Eref{tint} vanishes for all $\tau$, $\text{E}_{0}$, $\text{E}_{1}$, $\text{E}_{2}$ and $\text{E}_{3}$ if the Born rule holds:
\begin{align}
\label{int2}
\text{I}(\tau)&=\text{P}_{111}(\tau)-\text{P}_{110}(\tau)-\text{P}_{101}(\tau)-\text{P}_{011}(\tau)+\text{P}_{100}(\tau)+\text{P}_{010}(\tau)+\text{P}_{001}(\tau)-\text{P}_{000}(\tau) \nonumber \\
&=\frac{1}{3}+\frac{2}{9}[(\cos(\triangle_{2}-\triangle_{1})+\cos(\triangle_{3}-\triangle_{1})+\cos(\triangle_{3}-\triangle_{2}))\tau]-\frac{2}{9}[1+\cos(\triangle_{2}-\triangle_{1})\tau] \nonumber \\
&-\frac{2}{9}[1+\cos(\triangle_{2}-\triangle_{1})\tau]-\frac{2}{9}[1+\cos(\triangle_{2}-\triangle_{1})\tau]+\frac{1}{9}+\frac{1}{9}+\frac{1}{9}-0 = 0.
\end{align}
\section{EXPERIMENTAL SETUP}
\label{setup}
An ensemble of molecules with at least two spin-1/2s in LSNMR is suitable for the experiment. Spin-1/2 nuclei possess superpositions of up and down states under a static magnetic field and act as tiny bar magnets. In LSNMR quantum computing, an artibrary single qubit gate is implemented by applying electromagnetic pulses oscillating at radio frequency (RF pulses) along the plane perpendicular to the axis of the external magnetic field. Two qubit gates are achieved by modulating the evolution under \ql J-coupling"~\cite{levitt}. These single and two qubit gates enable universal control in LSNMR quantum computing. A large number of identical molecules precesses around the static magnetic field axis and forms a detectable magnetic signal allowing one to make a measurement.

\begin{figure}[!]
\centering
\epsfig{file=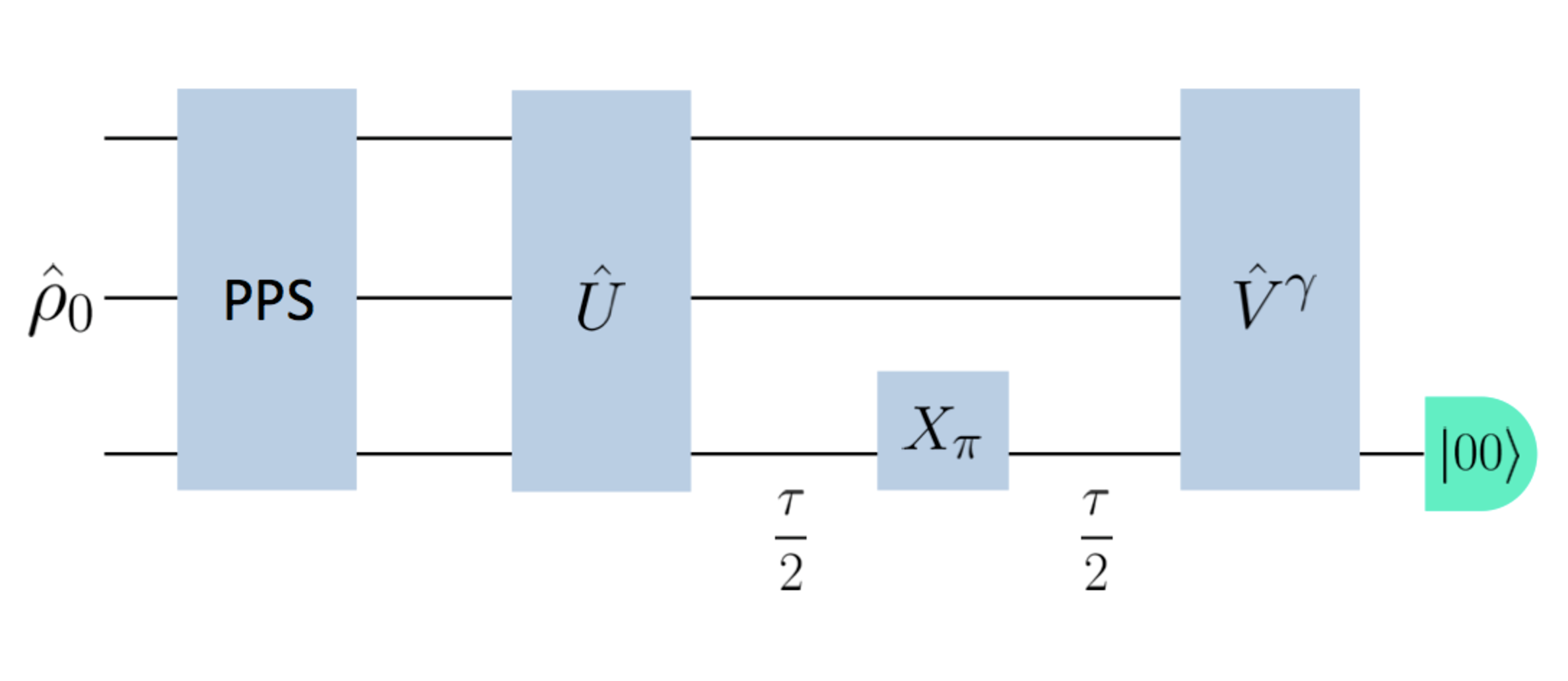,width=0.4\textwidth,angle=270}\vspace{-5mm}
\caption[Quantum circuit for the LSNMR three path experiment]{Quantum circuit for the three path experiment: The third (probe) qubit is used to read out the information encoded in the first two qubits. The deviation part \protect\cite{laflamme} of the initial thermal state is denoted as $\hat{\rho}_{0}$. The pseudo-pure state (PPS) $(\kket{00}\bbra{00}\otimes X)$ \protect\cite{laflamme} is prepared by an algorithm proposed in~\cite{knill2} with magnetic gradient pulses, and the whole preparation procedure is represented by PPS. An $X_{\pi}$ pulse applied to the third qubit at $\frac{\tau}{2}$ turns off unwanted interactions between the third read-out qubit and the two computation qubits. The unitary gate $\UU$ prepares $\kket{\psi^{111}}$, and the state evolves for $\tau$ into $\kket{\psi^{111}(\tau)}$. Applying unitary $\VV^{\gamma}$ and measuring the magnetization of the probe qubit conditional on the first two qubits being in the $\kket{00}$ basis give the probability $\text{P}_{\gamma}(\tau)$.}\label{qcircuit}
\end{figure}

The three path experiment with LSNMR can be illustrated as a quantum circuit in \Fref{qcircuit}.
Two qubits are used for encoding three paths, and a third (probe) qubit is added for read out. As a part of the initial state preparation, we perform \ql RF selection"~\cite{rfsel,colmthesis} in order to reduce the inhomogeneity of the RF field strength experienced by our liquid sample. Then a pseudo-pure state~\cite{laflamme} is prepared by an algorithm proposed in~\cite{knill2}. We use magnetic gradient pulses along the z-axis (the direction of static magnetic field) for labelling the coherence and decoding it to a pseudo-pure state. The output pseudo-pure state is $\alpha (\kket{00}\bbra{00}\otimes X)$ where $\alpha$ is the initial spin polarization. The unitary gate $\UU$ prepares $\kket{\psi^{111}}$, then the state undergoes free evolution for a time $\tau$. We apply an $X_{\pi}$ pulse at time $\frac{\tau}{2}$ after $\UU$ so as to \ql refocus"~\cite{laflamme} unwanted interactions between the third qubit and the two computation qubits. $\VV^{\gamma}$ transforms the amplitude of the interference from configuration $\gamma$ to the state $\kket{00}$ and yields the final state whose deviation part~\cite{laflamme} is $\hat{\rho}^{\gamma}_{f}=\alpha_{\gamma}'(\kket{00}\bbra{00}\otimes X)$, where $\alpha_{\gamma}'=\alpha P_{\gamma}(\tau)$. Then we measure the magnetization of the third qubit conditional on the first two qubits being in the $\kket{00}$ state. In other words, the signal is the overlap of the final state with the PPS, $\kket{00}\bbra{00}\otimes X$:
\begin{align}
M_{f}&=\text{Tr}[ \hat{\rho}^{\gamma}_{f}(\kket{00}\bbra{00}\otimes X)] \nonumber \\
&=\alpha_{\gamma}'\text{Tr}[(\kket{00}\bbra{00}\otimes X)(\kket{00}\bbra{00}\otimes X)]=\alpha_{\gamma}'.
\end{align}
Similarly, the initial PPS state magnetization is $M_{i}=\text{Tr}[\alpha(\kket{00}\bbra{00}\otimes X)(\kket{00}\bbra{00}\otimes X)]=\alpha$. Then
\begin{equation}
\label{msig}
\frac{M_{f}}{M_{i}}=\frac{\alpha_{\gamma}'}{\alpha}=P_{\gamma}(\tau).
\end{equation}

\begin{figure}[h]
\centering
\subfloat[]{\label{TTMSA}\epsfig{angle=270,width=0.5\textwidth,file=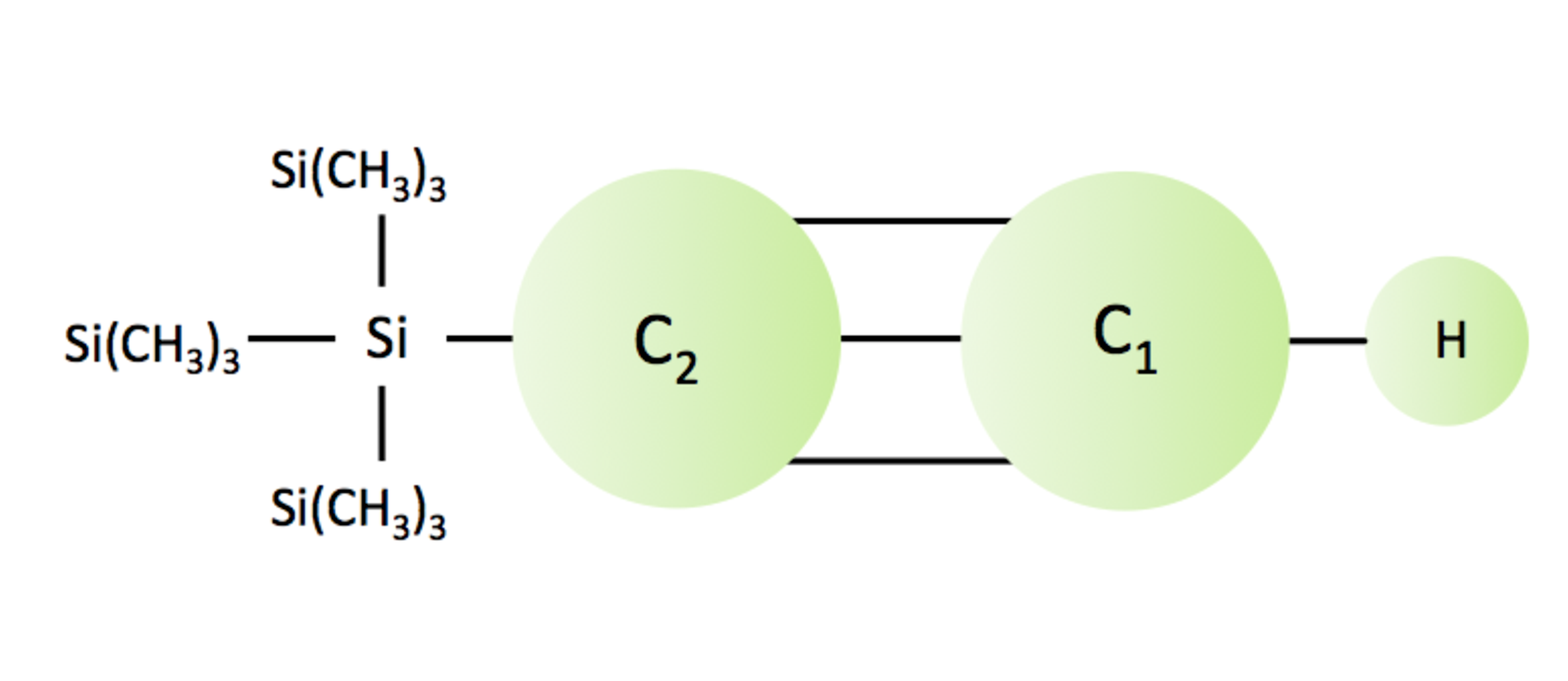}} \hspace{10mm}
\subfloat[]{\label{TTMSA_Ham}\epsfig{angle=270,width=0.4\textwidth,file=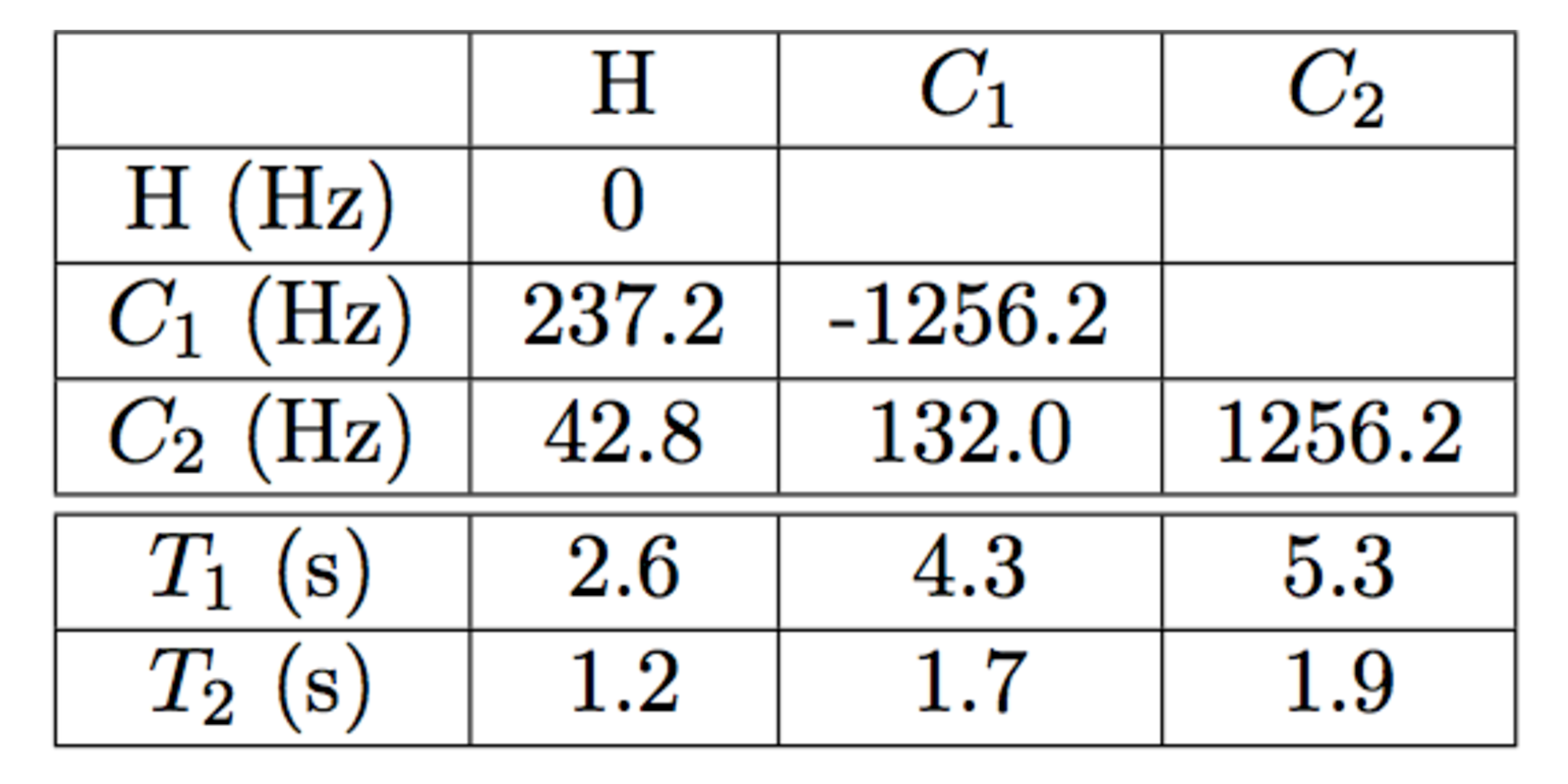}}
\caption{\label{TTMSA_all}(a) Schematic of a three qubit molecule used for the experiment (not in scale): A proton (\textsuperscript{1}H) and two carbons (\textsuperscript{13}C) are used for realizing qubits. (b) A table of natural Hamiltonian parameters (Hz), T1 and T2 (s): The diagonal elements give the chemical shifts with respect to the transmitter frequencies. The off-diagonal elements are the J-coupling constants.}
\end{figure}

The three path experiment was performed in LSNMR on a 700MHz Bruker Avance spectrometer at $293K$. A three qubit molecule was prepared from a sample of selectively labelled \textsuperscript{13}C tris(trimethylsilyl)silane-acetylene dissolved in deuterated chloroform (\Fref{TTMSA}). Natural Hamiltonian parameters that are relevant for the experiment are shown in \Fref{TTMSA_Ham}. Two \carbon 's are used to carry out the computation while the spectrum of \hydrogen\space is measured.
In the experiment, we used the Gradient Ascent Pulse Engineering (GRAPE)~\cite{colmthesis,grape} numerical optimization technique to find RF pulse shapes that implement the unitary evolutions $\UU$, $\VV^{\gamma}$, and the pseudo-pure state preparation with above $99.95\%$ Hilbert-Schmidt (HS) fidelity defined by
\begin{equation}\label{HSfidelity}
\Phi=\frac{\mid\text{Tr}[U^{\dagger}_{\text{app}}U_{\text{goal}}]\mid^{2}}{N^{2}}.
\end{equation}
We used a $20\mu$s square pulse for refocusing.

\begin{figure}[h]
\centering
\subfloat[Thermal State]{\label{TS}\epsfig{angle=270,width=0.4\textwidth,file=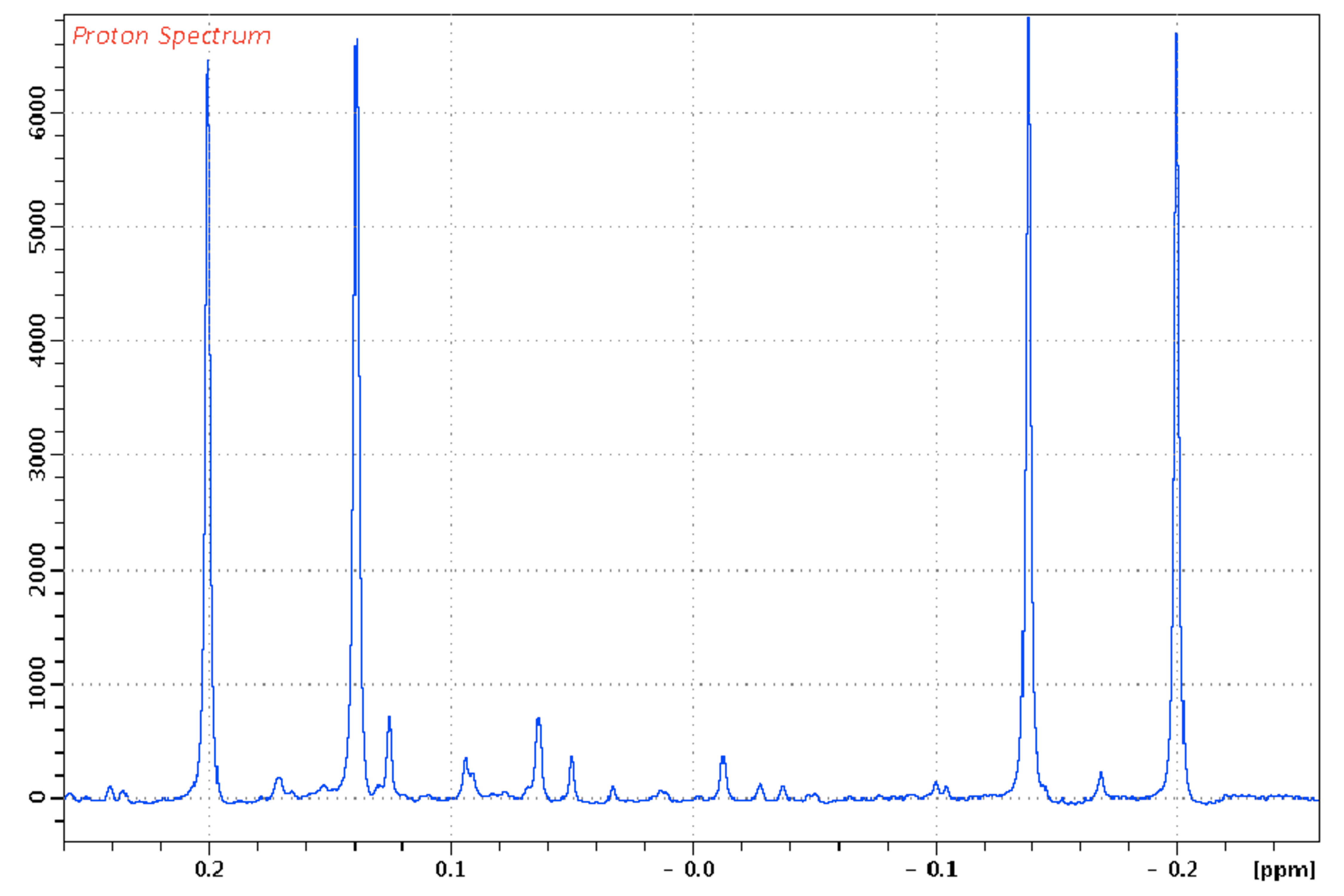}} 
\subfloat[Pseudo-Pure State]{\label{PPS}\epsfig{angle=270,width=0.4\textwidth,file=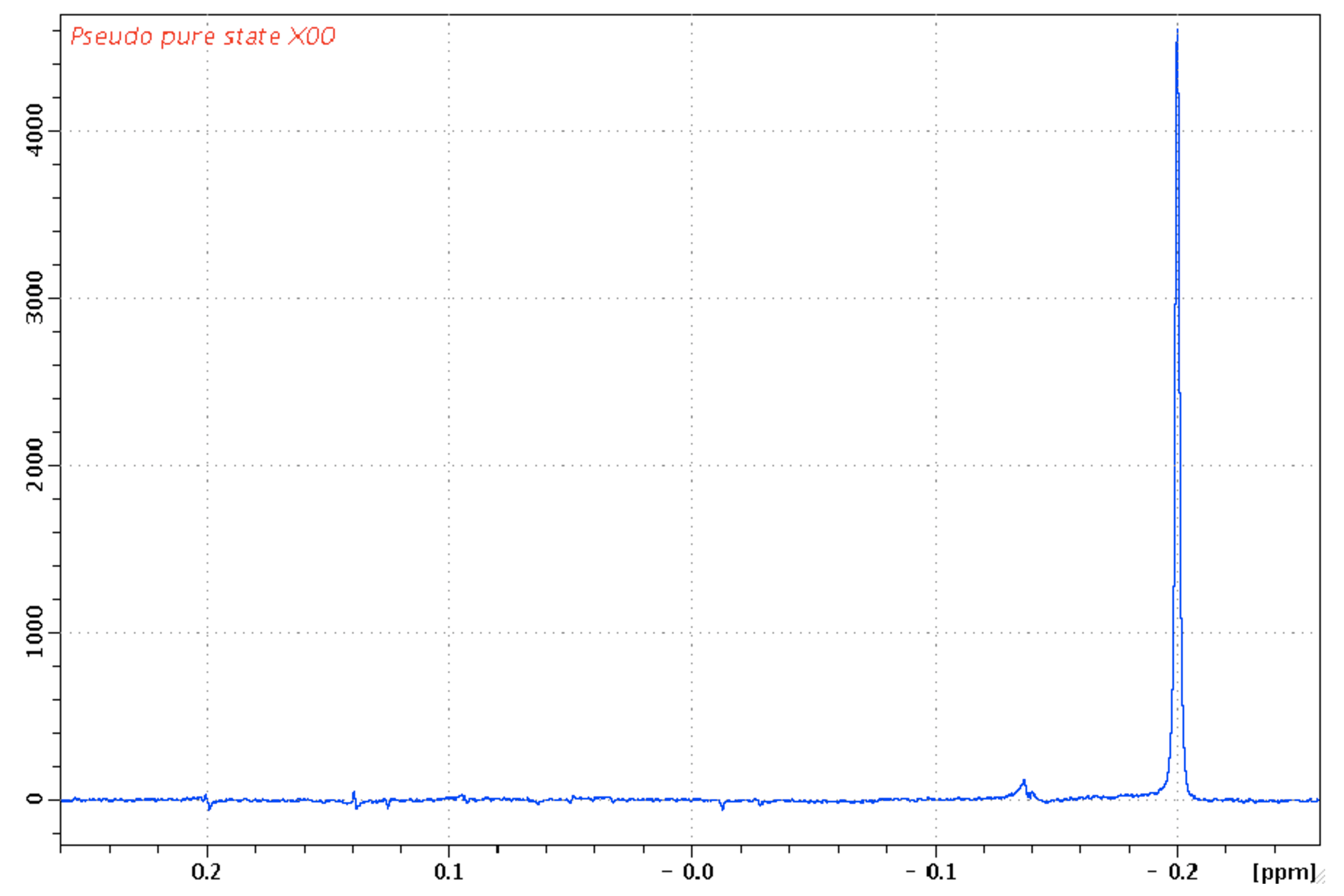}} \\
\subfloat[State after $\UU$]{\label{U}\epsfig{angle=270,width=0.4\textwidth,file=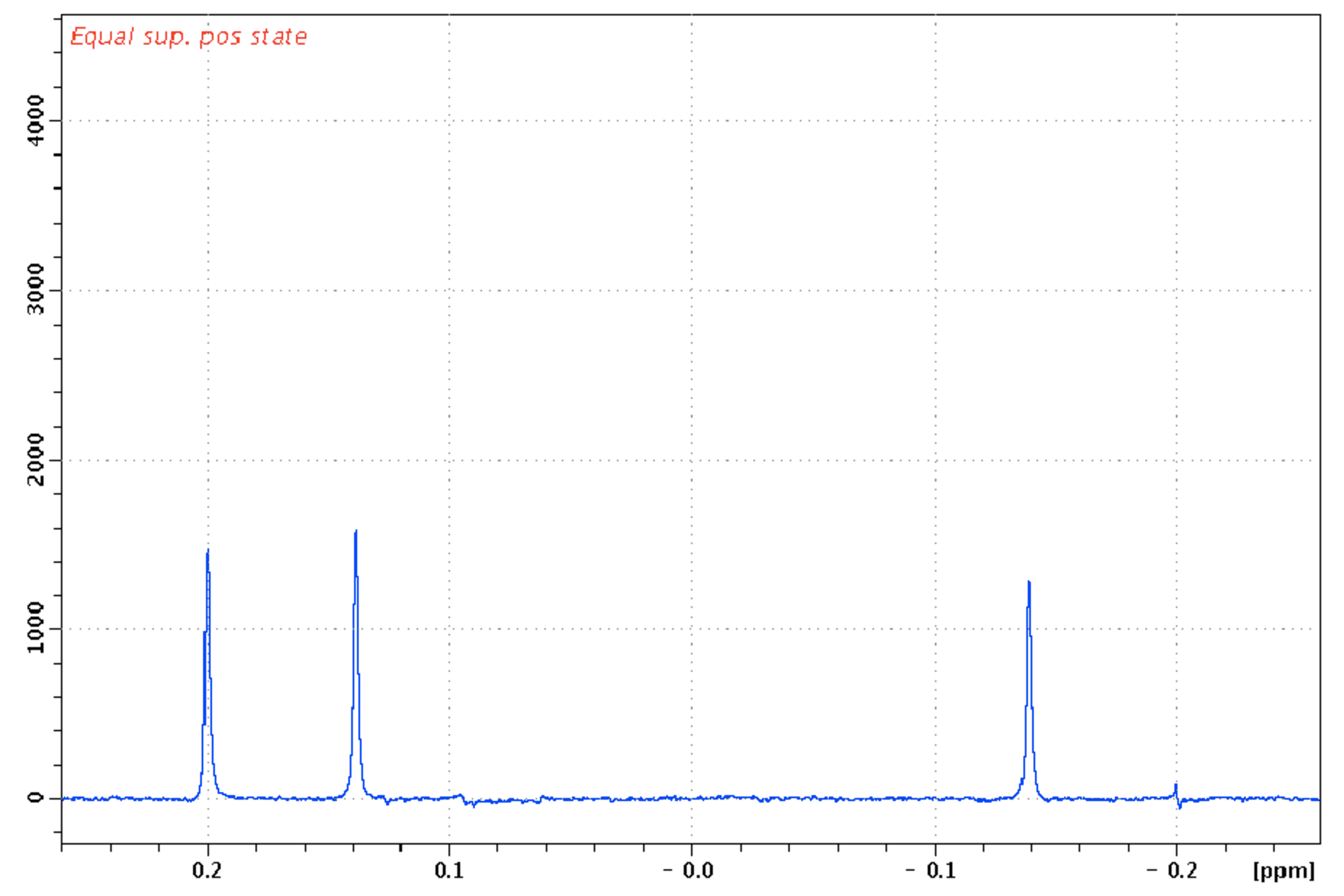}}
\subfloat[State after $\VV^{111}$]{\label{V111}\epsfig{angle=270,width=0.4\textwidth,file=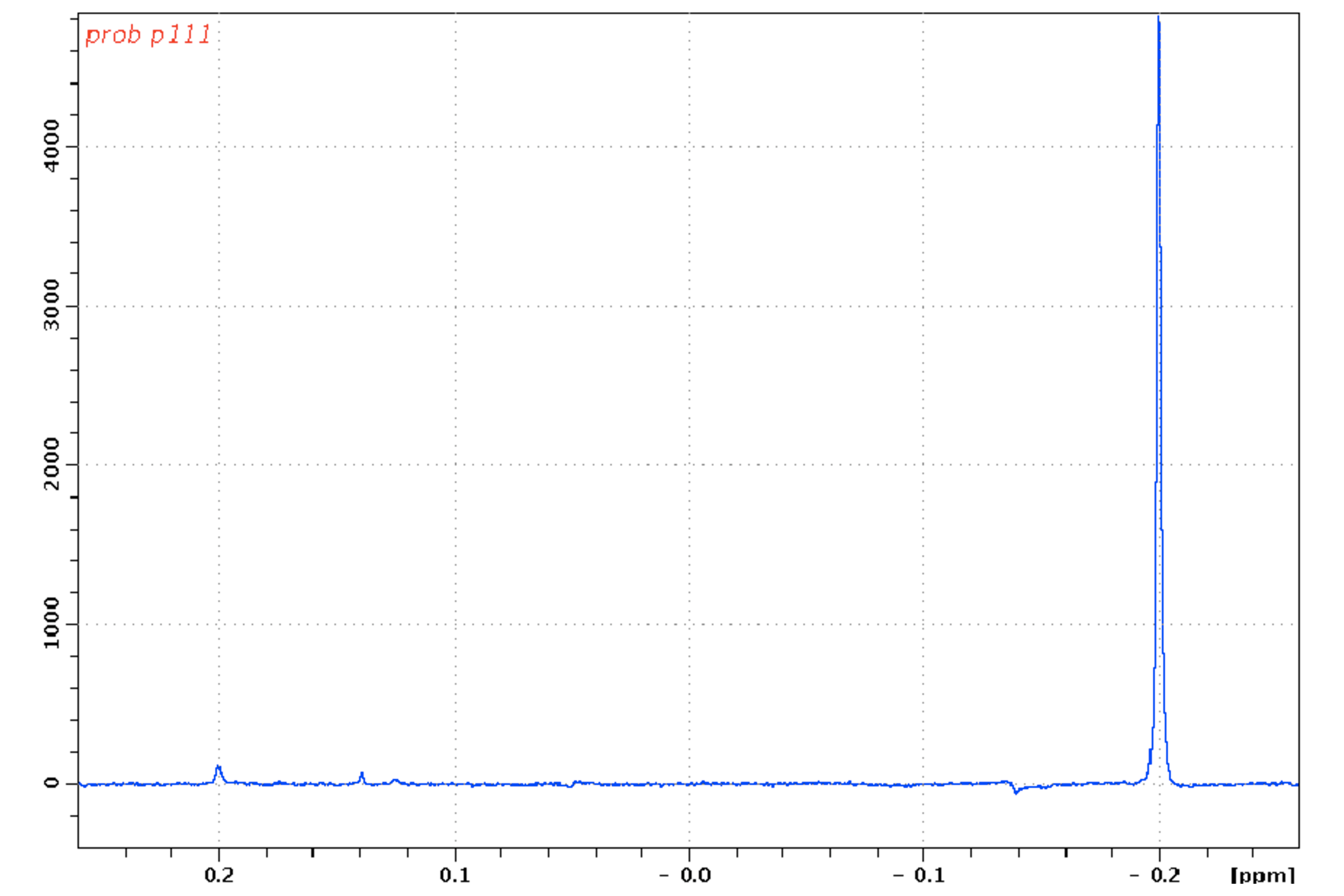}}

\caption[\textsuperscript{1}H in the Three Path Experiment]{\textsuperscript{1}H spectrum is used to gain information about the state of two \textsuperscript{13}C's. Vertical axis corresponds to magnetization amplitude and horizontal axis is frequency of spin precession. (a) is the initial thermal state spectrum of \textsuperscript{1}H, and the four peaks are due to four possible states of other two qubits. After applying PPS (\Fref{qcircuit}), we obtain a three qubits pseudo-pure state (b). (c) shows the equal superposition state $\kket{\psi^{111}}\bbra{\psi^{111}}$ of the two carbons. Finally, (d) is the spectrum we obtain after applying $\VV^{111}$.}\label{spectra}
\end{figure}

\Fref{spectra} illustrates an example of LSNMR spectra of \textsuperscript{1}H attained from an experiment for measuring $P_{\gamma}(\tau)$. For this particular example, $\gamma=111$ and $\tau=0$.

\section{RESULTS}
In the experiment, we evaluate the quantity
\begin{equation}
\label{rho}
\kappa=\frac{\text{I}(\tau)}{|\text{I}_{110}(\tau)|+|\text{I}_{101}(\tau)|+|\text{I}_{011}(\tau)|},
\end{equation}
where $\text{I}(\tau)$ is the three paths interference, and the denomenator is the sum of magnitudes of two paths interferences (e.g. $\text{I}_{110}(\tau)=\text{P}_{110}(\tau)-\text{P}_{100}(\tau)-\text{P}_{010}(\tau)+\text{P}_{000}(\tau)$). In this way, one can assure that the experiment is dealing with a quantum phenomenon inasmuch as the denomenator should vanish in classical case~\cite{sinha,sinha2}. Moreover, the calculation of such quantity is straight-forward in our experimental set up. 
As mentioned in the previous section, $\text{P}_{\gamma}(\tau)$ is obtained by normalizing the magnetization of the final state measured after unitary gate $\hat{V}^{\gamma}$ with that of the initial pseudo-pure state. Thus we run two experiments consecutively, first to measure the magnetization of the initial pseudo-pure state and second to measure the magnetization of the state acquired from the full quantum circuit (\Fref{qcircuit}). These two measurements are separated by $25s$ (about five times larger than T$_1$) in order for spins to re-thermalize.

\begin{figure}[!t]
\centering
\epsfig{file=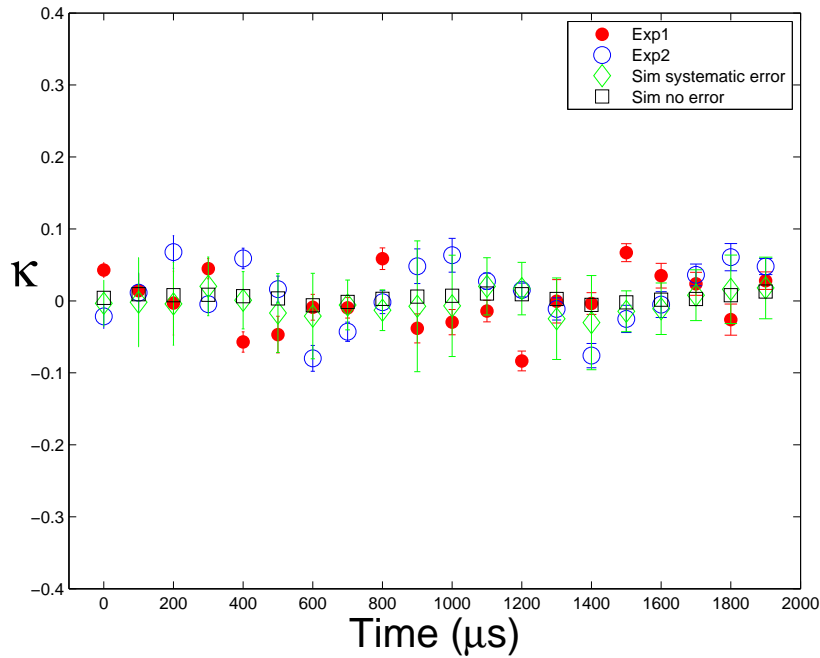,width=0.8\textwidth}
\caption[Plot of experiment results for $\kappa$]{Experimentally determined $\kappa$: The experiment is repeated ten times for each $\tau$, giving 200 data points in total. The red indicates the experimental outcomes and the size of the error bars represents standard deviation. We acquired from the experiment the weighted sample mean $\overline{\kappa}=0.007\pm 0.003$. The blue circles and error bars are obtained by repeating the experiment (again, 200 data in total) with the different measurement method explained in the following section. The results from the second method are $\overline{\kappa}=0.009\pm 0.003$. The black squares are obtained from the simulation which assumes the Born rule holds and perfect implementations of the GRAPE and hard pulses under the effect of T$_2$. Here $\overline{\kappa}=0.004\pm 0.001$, and its deviation from zero is due to the fildelity of GRAPE pulses and T$_2$. The green is the outcome of the simulation that takes systematic errors such as distortions in the implementation of GRAPE pulses and RF field fluctuations described in the next section into account. This simulation gives $\overline{\kappa}=-0.002\pm 0.001$. The systematic errors well explain the small deviation of experimentally determined $\kappa$ from zero; red and blue (experiment) overlap with green (systematic errors simulation) for most of the $\tau$'s, with less than one standard deviation away when there is no overlap.}
\label{rhoonly_new}
\end{figure}

We sample $\kappa$ for various $\tau$ from $0 \mu s$ to $1900 \mu s$ with discretization $\delta\tau=100 \mu s$. For each $\tau$, we repeat the experiment ten times providing 200 data in total. We obtained the weighted sample mean (WSM) $\overline{\kappa}=0.007\pm 0.003$. WSM is appropriate for the data analysis since the size of standard deviation varies for different $\tau$. The random error is the standard error of the WSM of $\kappa$. The results are shown in \Fref{rhoonly_new}. The red dots are the average of ten repetitions of the experiment and the size of the error bars indicate standard deviations of the average. The black circles represent simulation results. The simulation assumes that the GRAPE and hard pulses designed for the experiment are implemented with no error under the effect of T$_2$ and uses Born's rule to extract magnetization signal of the final state.

\section{Analysis of Possible Sources of Error}
\label{error}
In this section, we discuss possible sources of error of the experiment. As shown \Fref{TTMSA_Ham}, the
difference between the Larmor frequencies of $\text{C}_{1}$ and $\text{C}_{2}$ is an order of magnitude larger than the J-coupling and thus we are well into the weak coupling approximation. A simulation of the neglected strong coupling terms 
shows that on our time scale for $\tau$, they would contribute to $\overline{\kappa}\sim 10^{-16}$ and hence negligible.

Next, there are distortions in the implementation of shaped RF pulses~\cite{colmthesis}; the GRAPE pulses seen by the molecule in the LSNMR spectrometer do not exactly match to what we desire. There are two components to this deviation; random errors and a systematic portion that is primarily caused by limitations of the probe circuit design. The systematic imperfection can be rectified by placing a pick-up coil at the sample's place and closing a feedback loop to iteratively correct the RF pulse shapes~\cite{colmthesis,weinstein}. This method improves (yet, still not perfect) the closeness of the actual pulse to the desired pulse. Random fluctuations of the RF field are inevitable in the experiment. The RF variations for the \textsuperscript{1}H channel and the \textsuperscript{13}C channel are found to be $0.7\%$ and $0.2\%$, respectively. The RF selection process mentioned in \Sref{setup} is very sensitive to this RF field variations since it is designed to select a subset of the ensemble of spins at a specific nutation frequency~\cite{colmbm,colmthesis}. Thus the RF selection sequence in the presence of random RF fluctuations can introduce large fluctuations in the signal generated by pseudo-pure states. In other words, $M_i$ obtained from a reference pseudo-pure state deviates from $M_i$ of the following experiment in which the spin state goes through the full quantum circuit (\Fref{qcircuit}) and only $M_f$ is accessible. This leads to error in the probability calculation (\Eref{msig}), which in turn results in a non-zero mean value $\overline{\kappa}$. We prepared 100 pseudo-pure states following the RF selection consecutively and observed $0.95\%$ fluctuation of magnetic signal on average with the worst case being about $2\%$. This kind of fluctuation translates to $\overline{\kappa}\sim 10^{-3}\pm 10^{-4}$.

The green rhombi and error bars in \Fref{rhoonly_new} show the systematic errors due to distortions in the implementation of GRAPE pulses and the random fluctuation of RF field.

We performed another set of experiments with a different measurement method. Instead of taking two experiments to find the probabilities as described earlier, we opened the receiver at the end of the pseudo-pure state preparation for a short time, and opened the receiver again after $\hat{V}^{\gamma}$ so that the reference magnetization and the final magnetization are obtained from a single experiment. We hoped to see some improvement in this method by removing slow RF fluctuations between two experiments. Nevertheless, there was no significant improvement; we obtained $\overline{\kappa}=0.009\pm 0.003$. The results from this method is indicated as blue in \Fref{rhoonly_new}.

There are other possible sources of error such as transient effect from refocusing pulses due to their fast varying amplitude profile and disturbance of static field due to gradient pulses used for pseudo-pure state preparation. Furthermore, for the three qubit molecule TTMSA in LSNMR, the average error per gate is found to be $\sim 10^{-3}$ from randomized benchmarking in \cite{colmbm}. Such a gate error contributes to $\overline{\kappa}\sim 10^{-4}$.

\section{CONCLUSION}
The Born rule is one of the fundamental postulates of quantum mechanics and it predicts the absence of three way interference. We investigated the three way interference in order to test the consistency of the Born rule by performing LSNMR experiment using a three qubit molecule TTMSA (\Fref{TTMSA}). The quantum circuit for the experiment (\Fref{qcircuit}) was realized by composing GRAPE and hard pulses. We analyzed the quantity $\kappa$, the ratio of the three paths interference to two paths interference, and acquired $\overline{\kappa}\sim 10^{-3}\pm 10^{-3}$. Some of the major sources of experimental inaccuracy are listed in~\Sref{error}, and the simulation indicate that the small deviation of the experimental outcome from the Born rule is well explained by the systematic errors. Potential improvements of the experimental results include ensuring the pulse seen by the liquid-state sample is as close as possible to the ideal pulse and reducing RF inhomogeneity via enhenced probe design.

Finally, we conclude that the results of our experiment are consistent with Born's rule. Furthermore, we have demonstrated the capability of LSNMR QIP techniques for testing a fundamental postulate of quantum mechanics~\cite{PhysRevLett.107.130402,PhysRevLett.106.080401,contextuality,souzaNJP,souzaNJP2}.

\ack
D.P. would like to thank M. Ditty and J. Zhang for technical assistance with the LSNMR spectrometer and M. Laforest and C. Ryan for helpful discussions. This research is supported by CIFAR, Industry Canada and Quantum Works.
\section*{References}
\bibliographystyle{iopart-num}
\bibliography{reference}
\end{document}